\documentclass[pra,aps,groupaddress,showpacs,twocolumn]{revtex4}
\usepackage{epsfig,amsmath,graphicx,amssymb,overpic}
\def\be{\begin{equation}}
\def\ee{\end{equation}}
\def\bee{\begin{eqnarray}}
\def\ene{\end{eqnarray}}
\def\bes{\begin{subequations}}
\def\ees{\end{subequations}}

\begin{document}
\title{Analytical three-dimensional bright solitons and soliton-pairs
in Bose-Einstein condensates with time-space modulation }
\author{Zhenya Yan$^{1,2}$ and Chao Hang$^{2,3}$\vspace{0.05in}}
 \affiliation{$^1$Key Laboratory of Mathematics Mechanization, Institute of
Systems Science, AMSS, Chinese Academy of Sciences, Beijing
100080,
China \\
$^2$Centro de F\'isica Te\'orica e Computacional, Universidade de
Lisboa, Complexo Interdisciplinar, Lisboa
1649-003, Portugal \\
$^3$Department of Physics, East China Normal University, Shanghai
200062, China }



\begin{abstract}

We provide analytical three-dimensional bright multi-soliton
solutions to the (3+1)-dimensional Gross-Pitaevskii (GP) equation
with time and space-dependent potential, time-dependent
nonlinearity, and gain/loss. The zigzag propagation trace and the
breathing behavior of solitons are observed. Different shapes of
bright solitons and fascinating interactions between two solitons
can be achieved with different parameters. The obtained results  may
raise the possibility of relative experiments and potential
applications.

\end{abstract}
\vspace{-1in}
\maketitle


\baselineskip=12pt

\section{Introduction}

Solitons describe a class of fascinating
nonlinear wave propagation phenomena appearing as a result of
balance between nonlinearity and dispersion or diffraction
properties of the medium under nonlinear excitations, which leads
to undistorted propagation over extended distance \cite{Haus}. One of the most important
physically relevant realizations of solitons is provided by the
matter-wave solitons in Bose-Einstein condensed atomic gas
\cite{bec1}. Based on the successful experimental realization and
theoretical analysis of Bose-Einstein condensations (BECs) in
weakly interacting atomic gases \cite{bec1}, matter-wave
dark solitons~\cite{mws}, vortices~\cite{vor}, bright solitons~\cite{3dsol}, gap
solitons~\cite{gaps}, and soliton chains~\cite{chains} have been
observed and studied. These studies have stimulated a large amount of research
activities, which enable the extension of linear atom optics to
nonlinear atom optics \cite{bec2}.

The realization of higher-dimensional matter-wave solitons in BECs
is still a challengeable topic because those solutions are usually
unstable for (2+1)-D and (3+1)-D constant-coefficient nonlinear
Schr\"{o}dinger (NLS) equation due to the weak and strong collapse
\cite{Sulem}. However, different situations are observed in BECs
with temporally or spatially modulated parameters. Alteration of
atomic scattering length achieved by Feshbach resonance
\cite{feshbach} has been used to dynamically stabilize
higher-dimensional bright solitons \cite{stab1} while periodic
external potentials achieved by optical lattice has been used to
generate and control higher-dimensional gap solitons \cite{stab2}.
1D periodic wave solutions are also predicted in BECs with
time-space varying parameters \cite{1DNLS}. Moreover, the bright
solitons~\cite{3bec} and periodic wave solutions~\cite{yanc} were
obtained in spinor BECs governed by a system of three coupled
mean-field equations.

In this work, we present a detailed study on dynamics of {\it
analytical 3D bright matter-wave single solitons and
soliton-pairs} in BECs with time-space modulation. We note that 3D
periodic wave solutions have been studied in the generalized NLS
equation very recently \cite{belic, yank}. However, the authors
did not study the soliton pair solutions and their interaction
properties. By using the similarity~\cite{1DNLS,yank,simi} and
bilinear transformations~\cite{bi}, we can achieve different
shapes of bright solitons and fascinating interactions between two
solitons. In addition, the experimental possibilities for
observability are discussed and the stability of solitons is
illustrated numerically.

The paper is organized as follows. In the next section, the model
under study is introduced. In Sec. III, the methods for solving the
model equation are introduced. A relationship between the model and
a practical system is established. In Sec. IV, we give the
expressions of the bright solitons and the soliton-pairs. The
interactions between two solitons are also investigated. In the last
section, the case of the dark solitons is discussed and the outcomes
are summarized.


\section{The GP model}

The dynamics of a weakly interacting
Bose gas at zero temperature is well described by the (3+1)-D GP
model with time-space modulation~\cite{bec1}
 \bee
 \begin{array}{l} \label{gpe}
\displaystyle i\hbar\frac{\partial \Psi}{\partial t}\! =\! \left
[-\frac{\hbar^2}{2m}\nabla^2\!+ V_{\rm ext}(t,{\bf
r})+G(t)|\Psi|^2\right ]\!\!\Psi+i\,\Gamma(t)\Psi,\
\end{array} \vspace{-0.1in} \ene
where $\nabla=(\partial_x, \partial_y,
\partial_z)$, ${\bf r}=(x,y,z)$, $\Psi\equiv\Psi(t, {\bf r})$ denotes the order
parameter with $N=\int|\Psi|^2  d{\bf r}$ being the number of atoms
in the condensate, $G(t)=4\pi \hbar^2a_s(t) /m$ is the interaction
function with $a_s(t)$ being the $s$-wave scattering length
modulated by a Feshbach resonance, and $\Gamma(t)$ is the gain/loss
term, which is phenomenologically incorporated to account for the
interaction of atomic or thermal clouds. We note that the
dissipative dynamics originating from the interaction between the
radial and axial degrees of freedom has also been studied
recently~\cite{dissip}. Here the potential is chosen as a harmonic
trap $V_{\rm ext}(t,{\bf r})=(m/2)({\bf r}-{\bf e}(t))
\omega^2(t)({\bf r}-{\bf e}(t))$ with $\omega(t)={\rm
diag}(\omega_{x}(t), \omega_{y}(t), \omega_z(t))$ being a diagonal
 matrix of the  trap frequencies in three directions
and ${\bf e}(t)=$($e_1(t)$, $e_2(t)$, $e_3(t)$) corresponding to
its center.

Using the suitably scales and variables: ${\bf r}=a_{z}{\bf r}', \
t=\tilde{\omega}_{z}^{-1}t^{\prime}, \ {\bf e}(t)=a_{z}{\bf
e}'(t)$,\ $\Psi=\sqrt{N/a_{z}^3}\psi$, \
$a_{z}=[\hbar/(m\tilde{\omega}_{z})]^{1/2}$, and
$\tilde{\omega}_{z}=\int \omega_{z}(t) dt$, we arrive at the
dimensionless GP equation in the (3+1)-D space after dropping the
primes \bee \begin{array}{l}
 \displaystyle \label{dgp} i\frac{\partial\psi}{\partial t}
 =\left[-\frac{1}{2}\nabla^2+v(t,{\bf r})+
g(t)|\psi|^2\right]\psi+i\,\gamma(t)\psi, \end{array} \ene
 where
$g(t)=4\pi N^2a_s(t)/a_z^4$,
$\gamma(t)=\Gamma(t)/(\hbar\tilde{\omega}_z)$, and
\bee \label{v}
\begin{array}{l}
 \displaystyle v(t,{\bf r})=\frac{1}{2}({\bf r}-{\bf e}(t)) {\bf \alpha}^2(t)({\bf r}-{\bf e}(t)) \end{array} \ene
with $\alpha(t)={\rm diag}(\alpha_1(t), \alpha_2(t),
\alpha_3(t))=\tilde{\omega}_z^{-1}\omega(t)$. Eq. (\ref{dgp}) is
associated with $\delta\mathcal{L}/\delta \psi^*=0$ in which the
Lagrangian density can be written as
 \bee
 \begin{array}{l}
 \mathcal{L}=i(\psi\psi^*_t-\psi^*\psi_t)+|\nabla\psi|^2
-g(t)|\psi|^4 \vspace{0.1in}\cr \qquad \quad -2[v(t, {\bf
r})+i\gamma(t)]|\psi|^2.
 \end{array}
  \ene


\vspace{0.06in}

\section{Similarity solutions}

Here we focus on the spatially localized bright solitons and
soliton-paris for which $\lim_{|{\bf r}|\rightarrow
\infty}\psi(t,{\bf r}) =0$. Our first objective is to reduce Eq.
(\ref{dgp}) to the tractable NLS equation \bee \label{psi}
\begin{array}{l}
\displaystyle
i\frac{\partial\Phi(\tau,\xi)}{\partial\tau}
 =-\frac{1}{2}\frac{\partial^2\Phi(\tau,\xi)}{\partial\xi^2}+\mathcal{G}|\Phi(\tau,
\xi)|^2\Phi(\tau, \xi) \label{NLS}
\end{array} \ene using a proper similarity transformation, where
$ \tau\equiv\tau(t)$ and $\xi\equiv \xi(t,{\bf r})$  are both the
unknown variables, and $\mathcal{G}$ is a constant. We explore the
attractive nonlinearity, i.e. $\mathcal{G}=-1$, resulting in the
bright multi-soliton solutions. The case $\mathcal{G}=1$ resulting
in the dark multi-soliton solutions does not pose new challenges
and will be discussed in the last section. Using the similarity
transformation~\cite{1DNLS,yank,simi}
 \bee \label{Tr}
 \psi(t,{\bf r})=\rho(t)e^{i\varphi(t,{\bf r})}\Phi(\tau(t), \xi(t,{\bf r})),
\ene and requiring $\Phi(\tau(t),  \xi(t,{\bf r}))$ to satisfy Eq.
(\ref{NLS}) and $\psi(t,{\bf r})$ to be the solution of Eq.
(\ref{dgp}), we find a set of equations
\begin{subequations}
\label{sys}
\bee \label{sys1}
 &&
 \hspace{-0.2in}\begin{array}{l}
 \nabla^2\xi=0, \ \
  \xi_t+\nabla\xi\cdot \nabla\varphi=0, \ \   \tau_t=|\nabla\xi|^2,
  \end{array} \quad \\
\label{sys2}
 &&
\hspace{-0.2in} \begin{array}{l} \displaystyle v(t, {\bf
r})=-\frac{1}{2}|\nabla\varphi|^2-\varphi_t, \ \
g(t)=\mathcal{G}\rho^{-2}|\nabla\xi|^2, \end{array} \quad \\
\label{sys2}
 &&\hspace{-0.2in} \begin{array}{l}
 \displaystyle\gamma(t)=\frac{1}{2}\nabla^2\varphi+\rho_t/\rho.  \end{array} \quad
\ene
\end{subequations}

Here for the harmonic trapping potential $v(t, {\bf r})$ given by
Eq. (\ref{v}), after some algebra it follows from system
(\ref{sys}) that the similarity variables can be expressed as \bee
\label{variable}
\begin{array}{l}
 \displaystyle\xi(t,{\bf r})\!=\beta(t) \cdot {\bf r}\!+\! \int^t_{0}\beta(s)
\cdot \sigma(s)ds, \vspace{0.1in}\cr
 \displaystyle \tau(t)\!=\int^t_{0}
|\beta(s)|^2ds,
\end{array}
\ene where  $\beta(t)=(\beta_1(t), \beta_2(t), \beta_3(t))$ denotes
the vector of the inverse spatial widths of the localized solutions
along $x, y, z$ directions, respectively, and
$\sigma(t)=(\sigma_1(t), \sigma_2(t), \sigma_3(t))$ with
$\sigma_j=\beta_j\int^t_{0}e_j\alpha_j^2\beta_j^{-1}dt$ relating to
the velocity of the solitons. Moreover the nontrivial phase has the
quadratic form \bee \label{phase}
\begin{array}{l}
 \displaystyle\varphi(t,{\bf r})=-\frac{1}{2}{\bf r} A(t) {\bf r}+\sigma\cdot
{\bf r}-\frac{1}{2}\int^t_{0}({\bf e}\alpha^2{\bf
e}+|\sigma|^2)dt,
\end{array}
\ene where $A(t)={\rm diag}(\dot{\beta}_1/\beta_1,
\dot{\beta}_2/\beta_2, \dot{\beta}_3/\beta_3)$. The additional
relations between $\alpha_j(t)$ and $\beta_j(t)=1/\nu_j(t)$ result
in the Mathieu equations
 \bee
  \label{condition} \ddot{\nu}_j(t)+\alpha_j^2(t)\nu_j(t)=0,
  \quad (j=1,2,3).
  \ene
Finally, the function $\rho(t)$ modulating the amplitude of solution
$\psi$ and nonlinearity $g(t)$ can be also found by
 \bee
 \label{non}
 \begin{array}{l}
 \rho(t)=\displaystyle\rho_0[\beta_1(t)\beta_2(t)\beta_3(t)]^{1/2}
 \exp\left[\int_{0}^t\gamma(s)ds\right], \vspace{0.1in}\cr
 g(t)=-\rho^{-2}(t)|\beta(t)|^2,
 \end{array}
  \ene
which depend on both $\beta_j(t)$ and gain/loss coefficient
$\gamma(t)$ with $\rho_0$ being a non-zero parameter. Note that
for the given $\beta_j(t)$, the nonlinearity  $g(t)$ must {\it
attenuate} ({\it grow}) exponentially in the {\it gain} ({\it
loss})  medium $\gamma(t)>0 \ (<0)$.

For the given $\alpha_j(t)$, one can, in principle, obtain
corresponding $\beta_j(t)$ (or equivalently for the given
$\beta_j(t)$ one can obtain $\alpha_j(t)$) based on
Eq.(\ref{condition}). Furthermore, the bright $N$-soliton
solutions of Eq.(\ref{NLS}) can be obtained using the bilinear
transformation~\cite{bi}: $\Phi_N=P^{(N)}(\tau, \xi)/Q^{(N)}(\tau,
\xi)$. Here, $P^{(N)}$ and $Q^{(N)}$ satisfy
$(iD_{\tau}+1/2D_{\xi}^2)P^{(N)}\cdot Q^{(N)}=0$ and $D_{\xi}^2
Q^{(N)}\cdot Q^{(N)}=2|P^{(N)}|^2$ with $D_t$ and $D_{\xi}$ being
the bilinear operators and
$P^{(N)}=\sum_{j=1}^N\epsilon^{2j-1}P_{2j-1}(\tau, \xi)$ and
$Q^{(N)}=1+\sum_{j=1}^N\epsilon^{2j}Q_{2j}(\tau, \xi)$. Thus, by
choosing $\beta_j(t)$ and $\gamma(t)$, we can generate $v(t,{\bf
r})$ and $g(t)$ for which the generic bright $N$-soliton solutions
of Eq. (\ref{dgp}) can be found from Eq. (\ref{NLS}) on the basis
of Eq. (\ref{Tr}).  We will use this analytical result to
construct the exact bright $N$-soliton solutions with many
interesting nontrivial features.

For the convenience of analyzing different dynamical regimes
described by the given model, we specify the magnitude of main
physical parameters, which are feasible in experiments. We consider
a condensed sodium sample trapped in the state $|3S_{1/2}, F=1,
m_F=-1\rangle$, which has the scattering length $a_s=2.75$
nm~\cite{csys}. The other parameters can be taken as
$N=1.2\times10^6$ and $\tilde{\omega}_z=(2\pi)\times21$ Hz, which
leads to $a_z=4.55$ $\mu$m and $\sqrt{N/a_z^3}=1.13\times10^2$
$\mu$m$^{-3/2}$. To make sure the frequencies $\alpha_j(t)$ and
nonlinearity $g(t)$ are bounded for realistic cases, we choose
$\beta_j(t)$ and the gain/loss coefficient $\gamma(t)$ as the
periodic functions
 \bee \label{gamma}
 \beta(t)={\rm dn}(t, m){\bf b}, \quad
\gamma(t)=\gamma_0\, {\rm cn}(t,n), \ \ \ \gamma_0\in\mathbb{R}
 \ene
where ${\bf b}=(b_1, b_2, b_3)$ is a real constant vector
describing the inverse of the width of the potential and the
frequency, $m\in[0,\ 1)$ and $n\in[0,\ 1]$ are the modules of
Jacobi elliptic functions. It is easy to see that
$\gamma(t)=\gamma_0\,{\rm sech}(t)>0$ corresponds to the
dissipative case  when $n=1$ and $\gamma_0>0$ (we will focus on
this condition next). In practical systems, the modulations of
$\beta_j(t)$, $g(t)$ and $\gamma(t)$ depend on the use of the
optical lattice and Feshbach-resonance techniques, i.e. we can
achieve $\Gamma(t)$ and $a_s(t)$ by exerting particular
time-dependent optical field and magnetic field.

It follows from Eqs. (\ref{condition}) and (\ref{gamma}) that
$\alpha_j(t)$ is given by
 \bee
  \label{alpha}
  \alpha_j^2(t)=m^2\left[(2-m^2)\,{\rm sd}^2(t,m)-{\rm nd}^2(t, m)\right]. \ene
 Figure  1 shows the curves of $\alpha_j(t)$, $g(t)$, and
$\gamma(t)$ vs $t$. For simplicity, we take $b_j=1$, i.e.
$\alpha_1(t)=\alpha_2(t)=\alpha_3(t)$ corresponding to the isotropic
potential, and consider that the center of the potential locates at
the origin ($e_j=0$). We can also change $b_j$ to get an anisotropic
potential and use nonzero $e_j$ to obtain moving bright solitons as
those discussed in 1D case \cite{1DNLS}.

\begin{figure}
\begin{center}\hspace{-0.17in}{\scalebox{0.44}[0.30]{\includegraphics{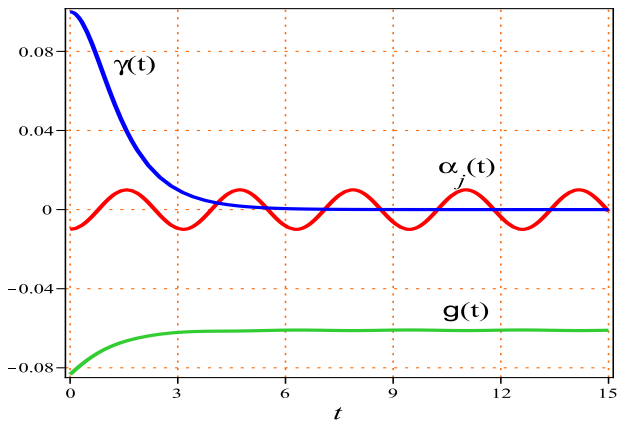}}}
\end{center}
\vspace{-0.25in}\caption{(color online). Curves of $\alpha_j(t)$,
$g(t)$, and $\gamma(t)$ given by Eqs. (\ref{gamma}) and
(\ref{alpha}) vs $t$ for $\rho_0=6.0$, $\gamma_0=m=0.1$, and
$b_j=n=1.0$ ($j=1$, 2, 3). }
\end{figure}

\section{Bright solitons and soliton pairs}

Based on the discussions in the previous section, we arrive at the
fundamental 3D time-varying bright solitons \vspace{-0.08in} \bee
 \label{one}
 \begin{array}{ll}
 \psi_1(t,{\bf r})
     \! =  r_1\rho(t)\, {\rm
    sech}\big[r_1(\xi(t,{\bf r})\!-\! s_1\tau(t))\!-\!
    \ln|2r_1|\big]e^{i\theta},\
 \vspace{-0.05in}
   \end{array}
\ene where $\theta=s_1\xi(t, {\bf
r})+(r_1^2-s_1^2)/2\tau(t)+\varphi(t,{\bf r})$ with $r_1, s_1\in
\mathbb{R}$, and $\rho(t)$, $\varphi(t,{\bf r})$, $\xi(t,{\bf
r})$, and $\tau(t)$ are given by Eqs. (\ref{variable}),
(\ref{phase}) and (\ref{non}).

Figure 2 exhibits the dynamics of the time-varying bright soliton
(\ref{one}). A breathing behavior is also evident, which can be
managed by $\beta_j(t)$, $\gamma_j(t)$, and $\rho(t)$. For the
case $m=0$, we have $\beta_j=b_j$ and $\alpha_j=0$, in which the
travelling-wave bright soliton is obtained. In experiments, it can
be simply realized for the zero linear potential. The bright
soliton propagates in a zigzag trace for $m=0.1$ [see Fig.2(a)].
An important feature is that while $m\rightarrow1$ ($\neq1$)
resulting in the larger period of $\beta_j(t)$ given by
Eq.(\ref{gamma}), the amplitude of the soliton close to the
corners attenuates rapidly so that a soliton chain is generated
[see Fig.2(b)]. In experiments, it can be realized by taking
$\omega_x=\omega_y=\omega_z=m[(2-m^2){\rm sd}^2(t,m)-{\rm nd}^2(t,
m)]^{1/2}\tilde{\omega}_z$.

\begin{figure}
\begin{center}\hspace{-0.11in}{\scalebox{0.45}[0.45]{\includegraphics{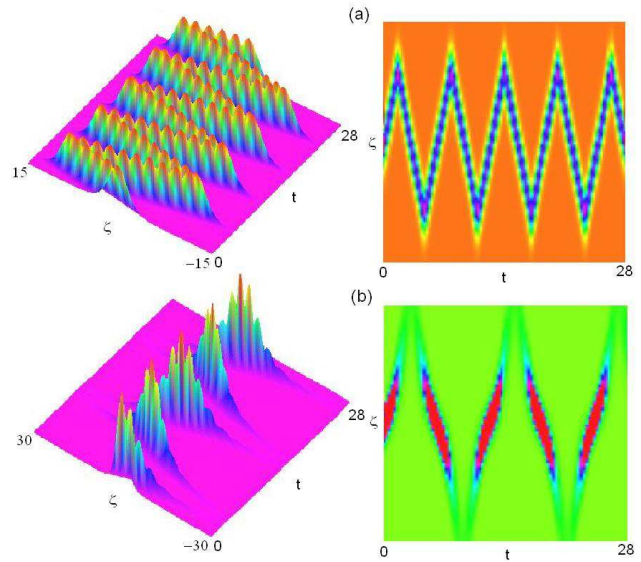}}}
\end{center}
\vspace{-0.4in}\caption{(color online). Propagations (left) and
contour plots (right) of density for the bright soliton
(\ref{one}) in $(t,\,\zeta\equiv {\bf b}\cdot {\bf r})$-space for
$r_1=0.5, \ s_1=\rho_0=2.0$, and $\gamma_0=0.01$. The others are
the same as Fig.1. (a) The breathing bright soliton propagating in
a zigzag trace for $m=0.1$. In the given sodium sample, the
maximum density and width of the soliton are about $1.28$
$\mu$m$^{-3}$ and $12.13$ $\mu$m. The period
is about $53.1$ ms. (b) The bright soliton chain for
$m=0.9$. The maximum density and width of the soliton are about
$1.28$ $\mu$m$^{-3}$ and $9.1$ $\mu$m. The period is about $91.0$
ms.}
\end{figure}

The interaction of the bright solitons plays an important role in
the study of BECs. Here, we will also study the interaction between
two bright solitons. The analytical 3D time-varying bright soliton
pairs read
 \bee
 \label{two}
 \begin{array}{l}
 \psi_2(t,{\bf r})=\rho(t)e^{i\varphi(t,{\bf r})} P(t, {\bf r})/Q(t, {\bf r}), \end{array}
 \ene
where $P(t, {\bf r})$ and $Q(t, {\bf r})$ can be expressed as the
series of exponential functions of $(t, {\bf r})$
  \bee
   \begin{array}{l}
   \displaystyle P(t, {\bf r})=\sum_{j=1}^2\delta_je^{\eta_j}+\sum_{j,k=1, j\not=k}^2\lambda_{jk}
  e^{\eta_j+\eta_j^*+\eta_k}, \vspace{0.1in} \cr
  \displaystyle Q(t, {\bf r})=1+\sum_{j,k=1}^2\Lambda_{jk}e^{\eta_j+\eta_k^*}+\Omega
 e^{\eta_1+\eta_2+\eta_1^*+\eta_2^*},
 \end{array}
   \ene
 with $\eta_j=\mu_j\xi(t, {\bf r})+\frac{i}{2}\mu_j^2\tau(t),\
\mu_j=r_j+is_j \ (r_j,  s_j, \delta_j\in \mathbb{R})$,
$\Lambda_{jk}=\delta_j\delta_k^{*}(\mu_j+\mu_k^*)^{-2},\
\lambda_{jk}=(\mu_k-\mu_j)\big[\delta_k\Lambda_{jk}(\mu_k+\mu_k^*)^{-1}
-\delta_j\Lambda_{kk}(\mu_j+\mu_k^*)^{-1}\big]\ (j\not=k)$, and $
\Omega=(|\delta_1\delta_2|)^{-1}|\mu_1-\mu_2|^2\big(\Lambda_{11}\Lambda_{22}
\sqrt{\Lambda_{12}\Lambda_{21}}
 -\Lambda_{12}\Lambda_{21}\sqrt{\Lambda_{11}\Lambda_{22}}\big)$.

The dynamics of the 3D time-varying bright two-soliton solutions
(\ref{two}) is exhibited in Figure 3. Under the different
parameters, we exhibit three cases for two weak zigzag solitons
without interaction [see Fig.3(a)], two strong zigzag solitons with
interaction [see Fig.3(b)], and strong-weak zigzag solitons with
interaction [see Fig.3(c)]. Notice that similar with the bright
solitons shown in Fig.2(b), for the case $m\rightarrow 1$
($m\neq1$), the amplitudes of the soliton-pairs close to the corners
will almost decrease to zero so that panel (a) will degenerate to
two parallel soliton chains while panels (b) and (c) will degenerate
to the $><$-shaped soliton chains. The experimental realization of
the dynamics regimes for the two-soliton solutions is similar with
that for the bright one-soliton solutions.

We stress that the important feature that distinguishes our
solutions from the reported in the literature
\cite{mws,vor,3dsol,chains} is the appearance of the time- and
space-dependent functions in both the phase and the amplitude and
which strongly affect the form and the behavior of bright solitons
and their interactions.

\begin{figure}
\begin{center}\hspace{-0.11in}{\scalebox{0.52}[0.50]{\includegraphics{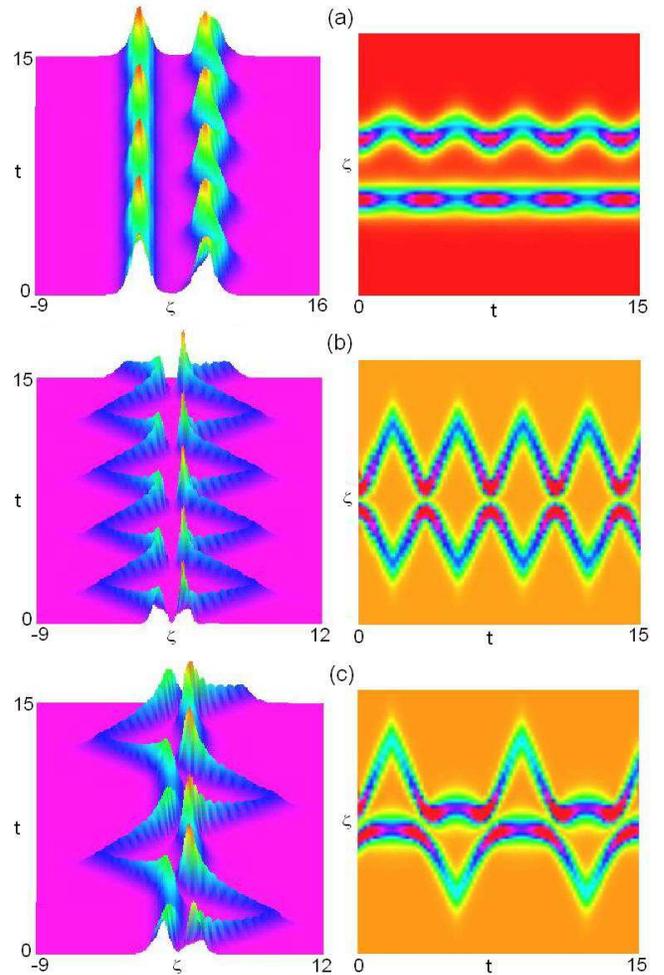}}}
\end{center}
\vspace{-0.25in} \caption{(color online). Propagations (left
) and contour plots (right) for collisions between bright two
solitons (\ref{two}) in $(t,\,\zeta\equiv {\bf b}\cdot {\bf
r})$-space for $\gamma_0=0.01, \ \delta_{1,2}=\rho_0=1.0$, and
$m=0.6$. The others  are the same as Fig.1.  (a) Two zigzag
solitons without interaction for $r_1=1.0, \ r_2=1.2$ and
$s_{1,2}=0$. In the given sodium sample, the maximum density and
width of the left (right) soliton are about $1.8$ $\mu$m$^{-3}$
and $7.6$ $\mu$m ($1.5$ $\mu$m$^{-3}$ and $9.1$ $\mu$m). (b) Two
strong zigzag solitons with interaction for
$r_{1,2}=s_1=-s_2=1.2$. The maximum density and width of the left
(right) soliton are about $6.2$ $\mu$m$^{-3}$ and $3.0$ $\mu$m
($17.5$ $\mu$m$^{-3}$ and $2.4$ $\mu$m) (c) Two strong-weak zigzag
solitons with interaction for $r_1=1.2,\ r_2=1.5,\ s_1=1.05$ and
$s_2=0$. The maximum density and width of the left (right) soliton
are about $6.2$ $\mu$m$^{-3}$ and $4.2$ $\mu$m ($13.1$
$\mu$m$^{-3}$ and $3.8$ $\mu$m). The period of the zigzag
oscillation is about $26.5$ ms in all panels. }
\end{figure}

In order to check the stability of the time-varying bright soliton
(\ref{one}), we make numerical simulations of Eq. (\ref{dgp}) with
the initial conditions given by Eq. (\ref{one}) and different values
of $m$. We find that the bright solitons are very stable for  $m$
being small (e.g. $m=0.1$) [see Fig.4(a)]. With the increase of $m$,
the bright solitons become unstable [see Fig.4(b)]. This is because
large $m$ results in stronger oscillations of $\beta_j(t)$, which
affect the coefficients of Eq. (\ref{dgp}) and the behavior of the
solutions.

\begin{figure}[!h]
\begin{center}
\hspace{-0.1in}
{\scalebox{0.46}[0.46]{\includegraphics{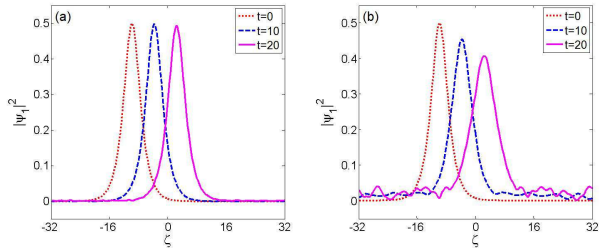}}}
\end{center}
\vspace{-0.25in}\caption{(color online). Numerical simulations of
bright soliton from Fig. 2 vs $\zeta$ at $t=0$, 10, and 20. The
initial conditions are given by Eq. (\ref{one}) with $s_1=0.1$,
$m=0.1$ in panel (a) and $m=0.9$ in panel (b). The other
parameters are the same as those used in Fig. 2.}
\end{figure}

\section{Discussions and Conclusions}

For completeness, we consider the repulsive nonlinearity in Eq.
(\ref{psi}), i.e. $\mathcal{G}=-1$. In this case, the equation
admits 3D dark soliton solutions in the form with a nontrial phase
 \bee
 \psi(t,{\bf r})=\big\{iv+k\tanh[k(\xi(t,{\bf r})-v
\tau(t))]\big\}e^{-i\mu \tau(t)}, \ene
where $\mu$ is the chemical
potential, $v=\sqrt{\mu-k^2}$, and $k$ is a free parameter
satisfying $k^2<\mu$.

In summary, we have analytically constructed the novel 3D
time-varying bright multi-soliton solutions for the (3+1)-D GP
equation with time-space modulation. We focus on the bounded
potential, nonlinearity, and gain/loss case to analyze the dynamics
of the breathing and the zigzag propagation trace of the obtained
solitons. Different shapes of the one-soliton solutions and the
fascinating interactions between soliton-pairs were achieved. The
stability of bright solitons have been checked numerically. The
method we present here can be extended to study the
higher-dimensional bright soliton solutions of other nonlinear
systems and their various interaction properties. The model
(\ref{dgp}) can also be extended to describe 3D nonlinear optical
media with varying coefficients \cite{belic} after the
transformation $z\leftrightarrow t$. The results we obtained may
raise the possibility of relative experiments and potential
applications.

\acknowledgments

The work of Z.Y. was supported by FCT under Grant No.
SFRH/BPD/41367/2007 and the NSFC60821002/F02. The work of C.H. was
supported by FCT under Grant No. SFRH/BPD/36385/2007.



\end{document}